\documentclass[twocolumn]{article}
\usepackage{a4}
\usepackage{times}

\begin{document}
\addtolength\headheight{-15pt}
\addtolength\textheight{25pt}
\addtolength\oddsidemargin{-25pt}
\addtolength\evensidemargin{-25pt}
\addtolength\textwidth{57pt}

\title{A shared write protected root filesystem
	for a cluster of network clients}

\author{Ignatios Souvatzis\thanks{
	Universit\"at Bonn, Institut f\"ur Informatik, R\"omerstr. 164,
	D-53117 Bonn, Germany; e-mail {\tt ignatios@cs.uni-bonn.de}}
	}

\date{}

\maketitle
\pagestyle{empty}
\thispagestyle{empty}


\renewcommand{\dbltopfraction}{0.4}
\setcounter{topnumber}{5}
\setcounter{totalnumber}{5}


\abstract{ 

A method to boot a cluster of diskless network clients from a single
write-protected NFS root file system is shown. The problems
encountered when first implementing the setup and their solution are
discussed. Finally, the setup is briefly compared to using a
kernel-embedded root file system.

}

\section{Introduction}

\begin{itemize}
\item Managing three diskless network clients can be done manually.
\item Manually managing ten is still possible, but tedious.
\item Manually managing hundreds is close to impossible.
\end{itemize}

When we got ten disk- (and head-) less network computers of a new type
that we wanted to use as computing nodes for a parallel virtual machine
for a practical course\cite{parlab}, we decided to set them up with a shared root
file system.

However, a BSD root file system has to be unique and writable for every
client machine for a couple of reasons, so that a single
writable shared root file system does not work.

As an alternate solution, we placed most writable directories onto
(virtual) memory file systems. The program area and configuration files
need only to be exported by the server for read only access. This way,
the system programs, system libraries, and the configuration are
protected against malicious users, even if they should gain root
privileges on the client machine.

This presentation elaborates on the problems we encountered and the
solutions we implemented, using stock NetBSD 1.5/1.6 as the client operating
system, with just a small script and a few configuration lines added.


\section{System Environment}

\subsection{File- and Bootserver}

SUN UltraSparc-10, originally running Solaris 2.6 (now Solaris 8),
located in a secure room. However, the problem and its solution do not
depend on the machine size and operating system, as long as the
clients' root file systems are mounted via something similar to NFS.

\subsection{Clients}

Originally 10 Digital Network Appliance Reference Design (DNARD)
machines, with 64 MB of RAM, no disk, no keyboard/monitor attached,
running NetBSD 1.4, later 1.5.3 (now 1.6),
located in a secure room. 

We believe that the solution could be applied, with some modification,
to other Unix-like operating systems, although certain features of
NetBSD did help a lot.

\subsection{Users}

The machines are to be used by 10 to 20 students for the duration
of a half-year course. They should be able to login from home. Even if we
successfully limit logins to them, we can't really trust them not to
break into the system when they can, or even more likely, by running 
some dangerous script given to them by evil people.


\section{Why a shared, read-only root?}

\subsection{Easier administration}

We want to install and configure on the server (or just one machine,
under special circumstances) and at most have to reboot the
(other) clients.

\subsection{Saving disk space}

NetBSD-1.5.3/arm32 or 1.6 needs about 20\,megabytes of disk space in the
traditional root file system (see figure \ref{du}).

\begin{figure}[htb]

\begin{verbatim}
server 196 # du -ks bin sbin etc
5795    bin
12760   sbin
606     etc
\end{verbatim}
\caption{NetBSD-1.5.3/arm32 root filesystem usage}
\label{du}

\end{figure}

While the total size (200 megabytes for 10 machines) does not look
like much today, even assuming 100 machines, disk space was more expensive when we started, and
{\em reliable} disk space, especially {\em backed up} disk space, has
not become cheaper as fast as unreliable (``desktop'') IDE disks.

\subsection{Server security}\label{sec}

The root filesystem {\em has to} be exported with (client-side) root
access rights. Leaving it writable allows a malicious client to
permanently change (at least) its own configuration.

(Securing user data against read-out or modification from a
manipulated client is beyond the scope of this work.)


\begin{figure*}
\input{paper.fstab.tex}
\caption{/etc/fstab}
\label{fstab}
\end{figure*}

\begin{figure*}
\input{paper.rcconf.tex}
\caption{/etc/rc.conf}
\label{rcconf}
\end{figure*}

\begin{figure*}
\input{paper.fsrun.tex}
\caption{The filesystems at run-time}
\label{fsrun}
\end{figure*}

\begin{figure*}[tp]
\input{paper.dhcp.tex}
\caption{DHCP server configuration file excerpt}
\label{dhcp}
\end{figure*}


\section{Problems found}

\subsection{Booting}

\begin{itemize}

\item The client has to learn its name and network address.

\item The client has to learn the name of its NFS swap file.

\item The client has to learn the name of the per-client filesystems
on the server.

\end{itemize}

\subsection{During operation}

Files on the root file system that are written on traditional
installations, or that are tied to a single machine, include:

\begin{itemize}

\item System log files, which are written to all the time.

\item {\tt *.pid} - files, written (at least) during startup of server
processes.

\item Sockets ({\tt /dev/log, /dev/printer,} ...), which are created by the
server processes that use them to communicate with their clients.

\item Device nodes (of terminal-like devices, including pty, mice,
keyboards) change their owner at each login and logout.

\item The ssh host key is stored in the root file system.

\item The package system database, normally {\tt /var/db/pkg}. If we
made {\tt /var} per-machine, we'd need a per-machine {\tt /usr/pkg},
too. Otherwise we would need to move the database to a shared location. 

\end{itemize}


\subsection{During shutdown}

{\tt /etc/nologin} is created by the first machine shutting down. It
will prevent login on the other clients.

This is fine for a coordinated shutdown of the whole cluster, but not
when booting a single client machine to test, say, a new kernel.

\section{Methods}

\subsection{Some problems aren't}

\begin{itemize}

\item As all our clients are equal in configuration and don't carry
any permanent state, there is no point to make them distinguishable in
a cryptographically secure way. So we just use the same set of host
keys for {sshd} on all the machines.

\item Network configuration for IPv4 is done by DHCP anyway (when
booting the machines), so we can also use it to learn the client name
(and a few other parameters). For IPv6, we're using stateless IPv6
autoconfiguration.

\item We configured syslog to send all logged events to the server. 

\end{itemize}

\subsection{DHCP server setup}

Here, nothing special is needed. The clients are set up with fixed
addresses and names assigned to them and learn the name of the kernel
to boot from and the location of the root file system (figure \ref{dhcp}).

\subsection{Sockets and {\tt *.pid} - files}\label{firstshroot}

During boot time, a small memory file system is created and
mounted on {\tt /var/run}. (BSD mfs stores its data in the address
space of the {\tt newfs\_mfs} process, so it's pageable\cite{mfs}). We added this
to {\tt /etc/fstab} (figure \ref{fstab}) and marked {\tt /var/run} as
a filesystem to be mounted very early in {\tt /etc/rc.conf} (figure
\ref{rcconf}). This is used for the following files:

\begin{itemize}

\item {\tt *.pid} files - often in {\tt /etc} - are created in {\tt
/var/run} by all daemons integrated into NetBSD or installed from the
package system.

\item {\tt /dev/log} was moved to {\tt /var/run/log} by changing the
syslogd code; this is the standard location in NetBSD nowadays.

\item {\tt /dev/printer} was moved to {\tt /var/run/printer}; this is
the standard location in NetBSD.

\end{itemize}

\subsection{Device nodes}

We create a memory file system for {\tt /dev}. {\tt /dev} on the
server only needs {\tt /dev/console} (for the benefit of {\tt /sbin/init}).

At boot time, we populate {\tt /dev} by running the MAKEDEV script,
which is installed in {\tt /sbin} in our setup. This takes about 20
seconds. Should we use slower machines, we could tune the amount
of devices created - currently, we run {\tt sh MAKEDEV all}.

The code to do this--as all special code needed--lives inside a small script 
called {\tt shroot} (figure \ref{shroot}).

\begin{em}
NetBSD-1.6 /sbin/init does all of this automatically, when no
/dev/console is found. (This was implemented to make CD-ROM and MS-DOS filesystem 
demonstration installations possible.) After upgrading, we were able to 
remove the {\tt /dev/console} on the exported root file system and remove
the lines in {\tt shroot} that handle {\tt /dev}. They are in comment lines
in figure \ref{shroot}.

\end{em}

\subsection{Swap, \tt /var}

During boot time we run {\tt /bin/hostname} to find out how we're
called - the kernel has learned it using DHCP. Using this name, we
synthesize the server-side names of the swap file and {\tt /var}
filesystem to mount.

\begin{figure*}[tp]
\input{paper.shroot.tex}
\caption{/etc/rc.d/shroot}
\label{shroot}
\end{figure*}

{\tt /var} is per-machine to allow per-machine spool files for
outgoing e-mail, printing and similar services.

\subsection{Remaining files written to {\tt /etc}}

Some of the files on {\tt /etc} can be configured not to be changed  (e.g., 
{\tt /etc/motd}). However, there are a few that can't be easily handled
without code and functionality change, like {\tt /etc/nologin}. 

As a simple catch-all to this problem, we create a small memory file
system and union-mount it over the NFS {\tt /etc}. This is handled in {\tt
/etc/rc.d/shroot}, too (fig. \ref{shroot}).

\subsection{Where to place the code?}

As mentioned already, we concentrated all special startup script code needed
in the {\tt shroot} script.

Obviously, this script has to run early in the boot process (before
device nodes, {\tt /var/run} etc. are accessed). Traditional {\tt /etc/rc.local}
is much too late. For NetBSD-1.4, we had hooked the equivalent script up in
{\tt /etc/netstart.local}; without that, we would have had to find a suitable
place in {\tt /etc/rc}, or among the zillions of SVR4 or Linux startup 
scripts.

The explicit startup script dependencies first implemented in NetBSD-1.5 \cite{rcconf}
made the task easy: the {\tt shroot} script is placed into the directory
{/etc/rc.d/} and states explicitly that it wants to be run after
the root file system is there, but before critical local file systems are
mounted (see figure \ref{shroot}).

This is important, because {\tt /var/run} has to be mounted after shroot 
mounts {\tt /var}!

Depending on applications, some subdirectories of {\tt /var} have to be
mounted from a shared NFS volume - e.g. {\tt /var/mail} or {\tt /var/games}.
This is handled normally in {\tt /etc/fstab} or using the automounter.

Figure \ref{fsrun} shows the run-time file system table. Note that
{\tt /usr} is embedded in the root file system! As it is never written
by the clients, there is no point to seperate it from root.

\subsection{Software Installation with the {\tt pkg}-system}

We are using a shared directory tree for the third-party packages
installed through the NetBSD package system. The package database is
moved inside {\tt /usr/pkg}, so that it is shared, too. 

During software installation, we give a single chosen client (which
temporarily gets a keyboard and monitor) write access to the server,
and revoke it afterwards.

The environment variable {\tt PKG\_DBDIR} has to be set to {\tt
/usr/pkg/libdata/pkgdb} for installation as well as any other use of
the {\tt pkg\_xxx} tools. Fortunately, this is all that is needed to
make the package system tools happy. {\tt pkg\_info} should be usable
by the students to find out what software is installed.

{\em However, from a security and performance viewpoint, it would be better
to have cross-install tools and to run them on the server.}


\section{Why no embedded root file system?}

An alternate method we've considered is to embed a root file system in
the client kernel. This leads to the same security benefits outlined
in section \ref{sec}. However, there are two drawbacks:

\begin{itemize}

\item An embedded filesystem is completely RAM based, non-pageable,
during execution. MFS, on the other side, is virtual memory based.

To make this work at all, the part of the root file system actually
inside the kernel has to be carefully tuned. The chosen method allows
to use a stock NetBSD installation, with only the {\tt shroot} script
added and some configuration files changed.

\item Both changing the kernel and changing some configuration file
require to embed the kernel file system anew into the kernel file
and reboot all clients.

This is inconvenient, especially when only the configuration of some
short-running component was to be changed.

The chosen setup allows to change all of {\tt /etc} on the server and have
it immediately available, if so desired.

\end{itemize}

\section{Summary}

A method to make the root file system for network clients read-only
and shared has been presented. Administration can mostly happen on the
server. Client break-ins would not affect the system files on the
server. The installation uses a mostly normal NetBSD-1.5.3 or -1.6
installation, with a single script added and some configuration files
in {\tt /etc} changed.


\end{document}